\documentclass[10pt,twocolumn,twoside,journal]{IEEEtran}
\IEEEoverridecommandlockouts
\usepackage{subfigure} 
\usepackage{graphicx}

\usepackage{amsmath,graphicx,amssymb,mathtools,bm}
\usepackage{subfigure}
\usepackage{hyperref}
\usepackage{cite}
\usepackage{amsmath,amssymb,amsfonts}
\usepackage{algorithmic}
\usepackage{graphicx}
\usepackage{textcomp}
\usepackage{xcolor}
\usepackage{verbatim}  
\usepackage{graphicx}  
\usepackage{bm}  
\usepackage{mathrsfs} 
\usepackage{algorithm} 
\usepackage{algorithmic} 
\usepackage{booktabs}
\usepackage{textcomp}  
\usepackage{multirow}  
\usepackage{lettrine}   
\usepackage{graphicx}  
\usepackage{color}  
\usepackage{amsmath}
\usepackage{amssymb}
\usepackage{stfloats}
\usepackage{geometry}
\usepackage{caption}
\usepackage{subfigure}

\usepackage{color}  

\UseRawInputEncoding

\def\BibTeX{{\rm B\kern-.05em{\sc i\kern-.025em b}\kern-.08em
    T\kern-.1667em\lower.7ex\hbox{E}\kern-.125emX}}
\begin{document}
	\newcommand{\tabincell}[2]{\begin{tabular}{@{}#1@{}}#2\end{tabular}}


\title{
	Enhancing Near-Field  Sensing and Communications with Sparse Arrays:  Potentials, Challenges, and Emerging Trends
}

\author{Songjie Yang, Wanting Lyu, Zhongpei Zhang, \IEEEmembership{Member,~IEEE} and Chau Yuen, \IEEEmembership{Fellow,~IEEE}


\thanks{Songjie Yang, Wanting Lyu, and Zhongpei Zhang are with the National Key Laboratory of Wireless Communications, University of Electronic Science and Technology of China, Chengdu 611731, China. 
	(e-mail:
	yangsongjie@std.uestc.edu.cn;
	lyuwanting@yeah.net;
	zhangzp@uestc.edu.cn).
	
	Chau Yuen is with the School of Electrical and Electronics Engineering, Nanyang Technological University (e-mail: chau.yuen@ntu.edu.sg).}}
 
\maketitle

\begin{abstract} 
As a promising technique, extremely large-scale (XL)-arrays offer potential solutions for overcoming the severe path loss in millimeter-wave (mmWave) and TeraHertz (THz) channels, crucial for enabling 6G. Nevertheless, XL-arrays introduce deviations in electromagnetic propagation compared to traditional arrays, fundamentally challenging the assumption with the planar-wave model. Instead, it ushers in the spherical-wave (SW) model to accurately represent the near-field propagation characteristics, significantly increasing signal processing complexity. Fortunately, the SW model shows remarkable benefits on sensing and communications (S\&C), e.g., improving communication multiplexing capability, spatial resolution, and degrees of freedom. In this context, this article first overviews hardware/algorithm challenges, fundamental potentials, promising applications of near-field S\&C enabled by XL-arrays. To overcome the limitations of existing XL-arrays with dense uniform array layouts and improve S\&C applications, we introduce sparse arrays (SAs).
Exploring their potential, we propose XL-SAs for mmWave/THz systems using multi-subarray designs. Finally, several applications, challenges and resarch directions are identified.

\end{abstract}
\begin{IEEEkeywords}
	Extremely large-scale arrays, near-field region, spherical-wave, sensing and communications, sparse arrays.
\end{IEEEkeywords} 
\section{Introduction}  
The sixth generation (6G) communication is expected to support various applications, such as ultra-high-definition video streaming, extended reality, autonomous vehicles, and IoT connectivity. To ensure stable and reliable high data rate services, breakthroughs in millimeter-wave (mmWave)/TeraHertz (THz) techniques with large bandwidths are urgently needed. Additionally, 6G introduces networked sensing, expanding beyond communication and enabling applications like localization, imaging, environmental monitoring, and gesture/activity recognition. The evolution to 6G holds immense potential for diverse applications, and advancements in technology will unlock transformative experiences and capabilities.

Nearly fifteen years ago, researchers started delving into the realm of spherical-wave (SW) propagation for mmWave Line-of-Sight (LoS) communications \cite{MIMO-LOS}. These investigations showcased the impressive multiplexing capability inherent in SW channels, presenting a promising opportunity to enhance high-frequency communication systems. The unique characteristics of SW propagation, such as the ability to support multiple data streams simultaneously, ignited interest in exploring their potential for communication technologies. Overall, these pioneering findings laid the foundation for further advancements in utilizing SW propagation to improve the capacity and efficiency of high-frequency communication systems, thereby paving the way for future developments in the field.

Recently, there has been significant attention towards extremely-large (XL)-arrays that prioritize SW propagation over planar-wave propagation in the near-field region. This region is bounded by the Fraunhofer distance, also referred to as the Rayleigh distance, which signifies the minimum distance required to maintain a phase difference of received signals across the array elements at a maximum of $\pi/8$. The amplified Fraunhofer distance resulting from the large aperture of XL-arrays necessitates acknowledgment within the typical communication coverage range, as its impact cannot be disregarded. Moreover, XL-arrays with their large number of antennas offer advantages such as improved signal quality, reduced interference, and enhanced spatial multiplexing capabilities. They not only enhance communication systems but also hold great promise for sensing applications \cite{Sensing1, Sensing2}. One significant aspect of this is  range estimation. By leveraging the near-field effect, narrowband range estimation becomes feasible, enabling applications such as object localization, target tracking, and proximity sensing.
 The combination of the SW propagation and XL-arrays opens up new possibilities for high-frequency communication and advanced sensing technologies.

	


XL-arrays using dense uniform arrays (DUAs) with half-wavelength spacing face two main limitations: high hardware costs and energy consumption due to numerous antennas, and restricted utilization of the SW property for medium-long distance communication, limited by the Fraunhofer bound. Overcoming these challenges is vital for advancing XL-arrays. Solutions include exploring alternative array configurations, array systhesis techniques, and adopting innovative approaches beyond traditional DUAs. This will unlock SW propagation's potential, empowering high-frequency S\&C applications.
Sparse arrays (SAs), as counterparts to DUAs, excel in array signal processing tasks, benefiting from their higher degrees of freedom (DoFs) and resolution. However, their full potential in communication systems, particularly in XL-arrays, remains untapped.

This article provides an overview of the hardware/algorithm challenges and fundamental potentials in near-field S\&C. It also explores the prospect of XL-arrays using SAs, referred to as XL-SAs, to uncover their untapped potential and inspire further research and innovation in this area.
 Specifically, we delve into SAs for XL-systems, covering: 1) an introduction of classical SAs, 2) a comparison between XL-SAs and XL-DUAs over some key features, 3) an integration of SAs into XL-systems, and 4)
 an outlook on the potential and applications of XL-SAs.
To support our findings, we conduct two simulations focusing on the channel rank and the Cramer-Rao bound (CRB) to showcase the potential of XL-SAs in near-field S\&C. Ultimately, we identify key challenges and corresponding potential research directions associated with near-field S\&C enhanced by XL-SAs.

\begin{figure*}
	\centering 
	\includegraphics[width=6.42in]{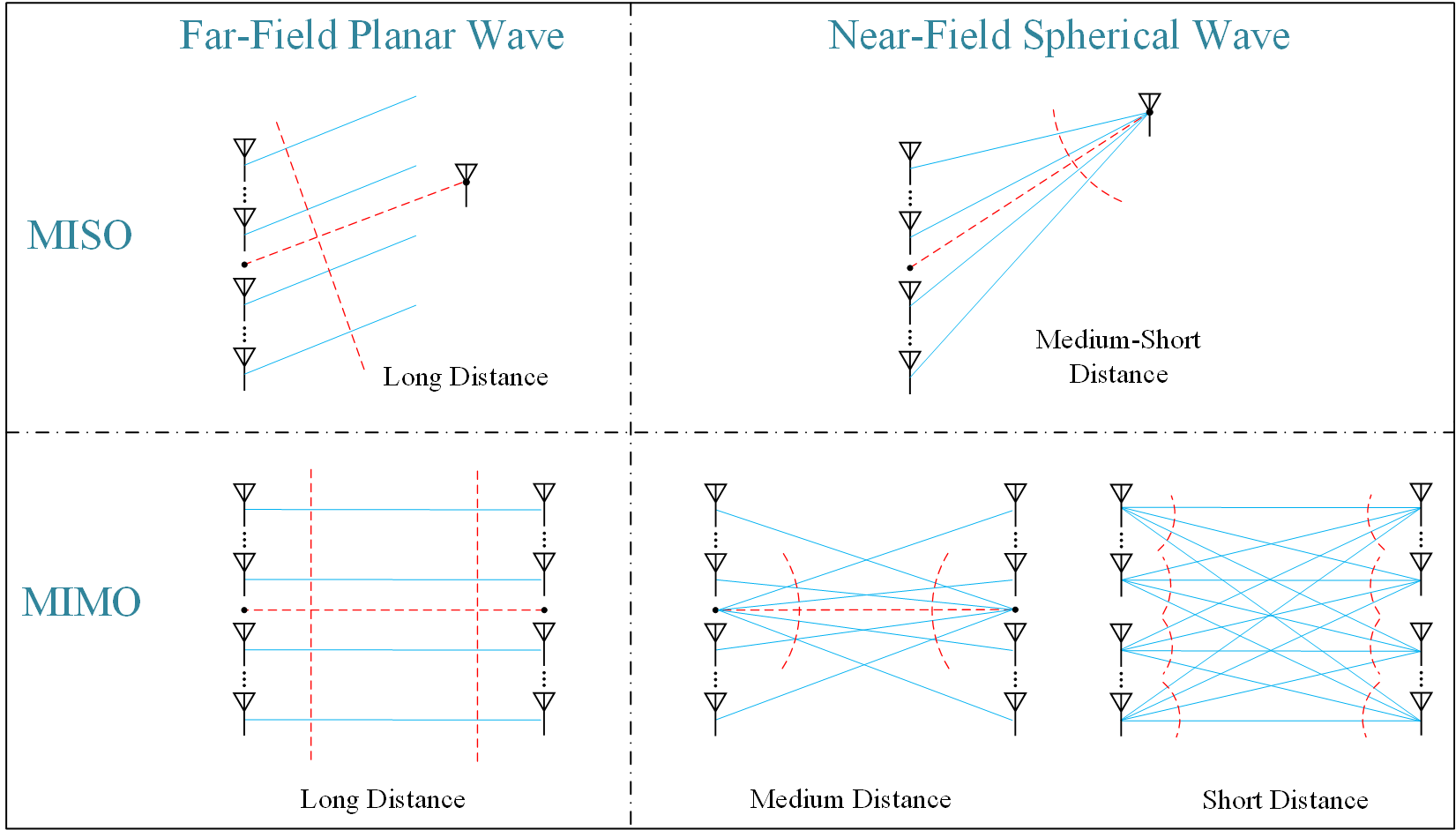}
	\caption{Planar-wave and spherical-wave propogation for MISO and MIMO cases}\label{SW} 
\end{figure*} 

\section{Near-Field XL-arrays: A Blessing in Disguise}\label{XL}
The original intent behind XL-arrays was to exploit the compact size advantage of high-frequency antennas while simultaneously increasing their quantity, thereby achieving significant beamforming gain. This innovative approach aimed to address the inherent challenges associated with high loss in high-frequency communication. 

However, the large number of antennas poses significant challenges for hardware systems, encompassing cost, complexity, and design. Moreover, the extensive array aperture introduces deviations in electromagnetic propagation compared to traditional arrays, thereby amplifying the complexity of signal processing. Nevertheless, challenges are often accompanied by opportunities. By harnessing the near-field effect, XL-arrays can unlock additional benefits beyond their substantial beamforming gain.

\subsection{Hardware Constraints}

Despite using low-complexity hybrid beamforming, XL-Multi-Input Multi-Output (MIMO) faces more hardware design challenges than massive MIMO in mmWave/THz systems.
\subsubsection{Hardware Cost}
 In hybrid beamforming architectures, each antenna necessitates vital components: at least one power amplifier/low-noise amplifier for signal transmission or reception and a phase shifter for directional control. As the number of antennas grows, the demand for phase shifters and power amplifiers/low-noise amplifiers increases linearly, leading to substantial hardware costs and higher energy consumption.

\subsubsection{Control Complexity}
 Phase errors arising from manufacturing tolerances and material imperfections need consideration. These random errors have a multiplicative impact on the beam pattern, necessitating algorithmic compensation \cite{HC}. However, the simultaneous management of numerous phase shifters in XL-arrays poses an even more challenging task.
 Moreover, practical power amplifiers introduce nonlinearity, causing distortions in the transmitted signal and interference with adjacent tones. Digital pre-distortion, as one efficient solution, mitigates this by pre-processing the signal based on captured power amplifier characteristics. However, the drawback of digital pre-distortion is its increasing computational complexity with antennas, and adopting ultra-large bandwidth may further limit its usage in XL-arrays. Array diagnosis is another challenge involving blocked antenna detection, identifying power loss or phase shifts due to blocking particles. In the near-field region, fast and reliable algorithms are challenging for continuous monitoring, identifying abnormalities, and ensuring efficient XL-system operation.
\subsubsection{Physical Design}
In practice, there are certain manufacturing challenges. 
Integrating numerous antenna elements within an array results in a significant escalation in the complexity of routing circuit components and antenna elements. Moreover, the heightened number of components concentrated in a confined space also elevates the cooling requirements for these circuits. As a consequence, the design and implementation of such XL-arrays demand meticulous attention to ensure efficient and reliable operation, as well as to manage potential thermal challenges associated with the increased circuit density.  
\begin{table*}
	\caption{Comparison of Arrays with Fixed Antenna Numbers and Maximum Apertures}
	\label{TAB}
	\centering
	\begin{tabular}{ |c|c|| c | c |c|c| }
		\hline
		\tabincell{c}{Two Cases}& Array Layout & Antenna Positions $[d]$ & \tabincell{c}{Number of \\ Antennas} &\tabincell{c}{Maximum \\ Aperture $[d]$} &DoFs \\ \hline\hline
		\multirow{8}{*}{\tabincell{c}{Fixed Number \\of Antennas}} & DUA & $\{0,1,2,3,4,5,6,7\}$ &8&8 &15  \\ \cline{2-6}
		& NA& $\{0,1,2,3,4,9,14,19\}$&8&20 &39\\
		\cline{2-6}
		& CA & $\{0,2,4,5,6,8,10,15\}$&8&16&27\\
		\cline{2-6}
		& NRA & $\{0,1,4,9,15,22,32,34\}$&8&35&57 \\
		\cline{2-6}
		&WSMS & $\{0,1,4,5,8,9,12,13\}$&8&14&21 \\
		\cline{2-6}
		&Proposed NMS &$\{0,1,2,3,4,5,10,11\}$&8&12&23 \\
		\cline{2-6}
		&Proposed CMS &$\{0,1,2,3,4,5,6,7\}$&8&8&15 \\
		\cline{2-6}
		&Proposed NRMS & $\{0,1,2,3,8,9,12,13\}$&8&14&27 \\
		
		\hline
		\multirow{8}{*}{\tabincell{c}{Fixed Maximum \\ Aperture}} & DUA & $\{0,1,2,3,4,5,\cdots,31,32,33,34,35\}$ &36&36 &71  \\ \cline{2-6}
		& NA& $\{0,1,2,3,4,5,11,17,23,29\}$&10&36&59 \\
		\cline{2-6}
		& CA & $\{0,4,5,8,10,12,15,16,20,25,30,35\}$&12&36 &59 \\
		\cline{2-6}
		& NRA &$\{0,1,4,9,15,22,32,34\}$&8&36&57\\
		\cline{2-6}
		&WSMS &$\{0,1,4,5,8,9,12,13,16,17,20,21,24,25,28,29,32,33\}$&18&36&51 \\
		\cline{2-6}
		&Proposed NMS & $\{0,1,2,3,4,5,6,7,14,15,22,23,30,31\}$&14&36&63\\
		\cline{2-6}
		&Proposed CMS & $\{0,1,4,5,8,9,10,11,12,13,16,17,20,21,30,31\}$&16&36&59 \\
		\cline{2-6}
		&Proposed NRMS & $\{0,1,2,3,8,9,20,21,24,25,34,35\}$&12&36&65 \\
		\hline

	\end{tabular}
\end{table*}
\subsection{Near-Field Signal Processing Challenges}

The near-field SW model involves complex mathematical expressions, particularly challenging for practical applications with multiple or complex sources. 
Fig. \ref{SW} displays five distinct scenarios of electromagnetic wave propagation. In the Multi-Input Single-Output (MISO) scenario or when sensing point sources, far-field planar waves result in antennas exhibiting uniform directional transmission with linear phase differences. On the other hand, near-field spherical waves, determined by distance and angle, create nonlinear phase differences among antennas. In the MIMO scenario or when sensing large aperture targets, near-field spherical waves demonstrate more complex properties compared to far-field planar waves. At long distances, both arrays approximate as point sources, forming plane waves. At medium distances, the arrays act as mirror images, forming a mirror-image MISO configuration. At short distances, each antenna acts as an individual point source, resulting in complete SW transmission.


 \subsubsection{SW Parameter Estimation}
 Classical estimation algorithms, like subspace methods and compressive sensing, face challenges with XL-arrays. Some subspace methods demand intensive computations due to eigenvalue decomposition, while compressive sensing encounters difficulties in sparse recovery due to the large-scale dicationary  resulting from the
 SW model's additional DoF. Moreover, near-field parameter tracking becomes more challenging compared to far-field tracking, as it involves rapid deviations with changes in transmitter-source distance.
 
 \subsubsection{MIMO-LoS Channel}
The MIMO-LoS channel has been extensively studied, with a focus on maximizing the spherical-wave channel capacity through optimized array aperture and antenna positions. On the other hand, the unique nature of the MIMO-LoS channel presents challenges for channel estimation, especially when the communication distance is short. In such a case, as illustrated in Fig. \ref{SW}, the channel cannot be divided into an outer product of two array response vectors. This aspect poses a significant challenge when attempting to estimate LoS and NLoS channel paths jointly. 
 
 \subsubsection{Array Synthesis}
Array synthesis is vital in array signal processing, aiming to optimize antenna excitation coefficients to meet criteria like beam directivity and bandwidth. It can also optimize antenna positions and their number for desired beam patterns. Traditional synthesis focuses on the far-field beam pattern, simpler than the near-field one, which becomes complex due to nonlinear terms in the array response.
 \subsubsection{Array Calibration}

Accurate beam pointing in an array antenna requires precise control of phase and amplitude, necessitating periodic calibration due to electronic component variations with temperature and time. Array calibration involves tasks like analyzing radiation patterns, calibrating beamforming, and assessing signal-to-noise ratio. However, measuring each antenna individually can be impractical due to spatial constraints with a large number of antennas. Furthermore, ensuring proper element excitation, phase synchronization, and mutual coupling compensation in an XL-array can be time-consuming and error-prone.

In addition to the mentioned challenges, the substantial increase in spatial dimensions will lead to extremely high computational complexity in XL-systems. This demands expensive computing resources and hinders real-time applications.

\subsection{Fundamental Potentials}  
Despite the challenges that exist in XL-arrays, there are several potentials that are worth exploring. 
Beyond the substantial beamforming gain, XL-arrays bring forth three fundamental potentials for S\&C.
\subsubsection{Multiplexing}
In mmWave MIMO-LoS communications, the SW channel has exhibited a high channel rank, enhancing multiplexing capability for simultaneous transmission of multiple data streams and indirectly improving capacity. 
\subsubsection{Spatial Resolution}
 XL-arrays showcase a formidable beamforming ability, providing an exceptionally directive beam that not only covers the angle but also extends to the range, resulting in high-resolution parameter estimation. 
\subsubsection{Range Sensing}

The commonly used range sensing method is time-of-arrival, which estimates delay for range calculation, limited by frequency resources for spatial resolution. However, using the SW model enables near-field range estimation in a narrowband system. Combining time-of-arrival and SW propagation shows potential for robust range estimation.

\subsection{Promising Applications}

Drawing from the listed potentials above, we can identify three promising applications that align with them. 
\subsubsection{mmWave/THz Reconfigurable Intelligent Surface (RIS)}
The mmWave/THz RIS encounters a low-rank channel challenge, resulting in weak multiplexing capability. Planar-wave-based RIS systems face issues in strong sparse scenarios, where only the LoS path exists between the tramister and the RIS, limiting simultaneous user service. To overcome these limitations, the SW channel emerges as a promising solution by increasing DoFs associated with the LoS path, making RIS-LoS an investigated approach \cite{RIS-LoS}. Moreover, 
\subsubsection{Spatial Division Multiplexing Access (SDMA)} Based on the second fundamental potential, XL-arrays show promise in offering high-resolution and two-dimensional features in the near-field region, facilitating user access in both angle and range dimensions, enabling SDMA. A similar concept was proposed in \cite{LDMA}, where location division multiplexing access employed user locations to distinguish users using a designed distance-angle codebook.
\subsubsection{Near-Field ISAC}
Since the SW property can not only enhance wireless communications but also find valuable applications in sensing, such as localization \cite{RIS-LOC}, it provides a significant opportunity for integrating S\&C in the near-field region. This dual functionality makes XL-arrays a promising technology with far-reaching implications.

\section{Enhancing Near-Field S\&C with XL-SAs}
To boost an array antenna's resolution or other attributes, the common approach is extending the array length. Yet, a smarter choice involves strategic element placement, like using SAs, even when the element count is fixed.
\begin{figure*}
	\centering 
	\includegraphics[width=6.42in]{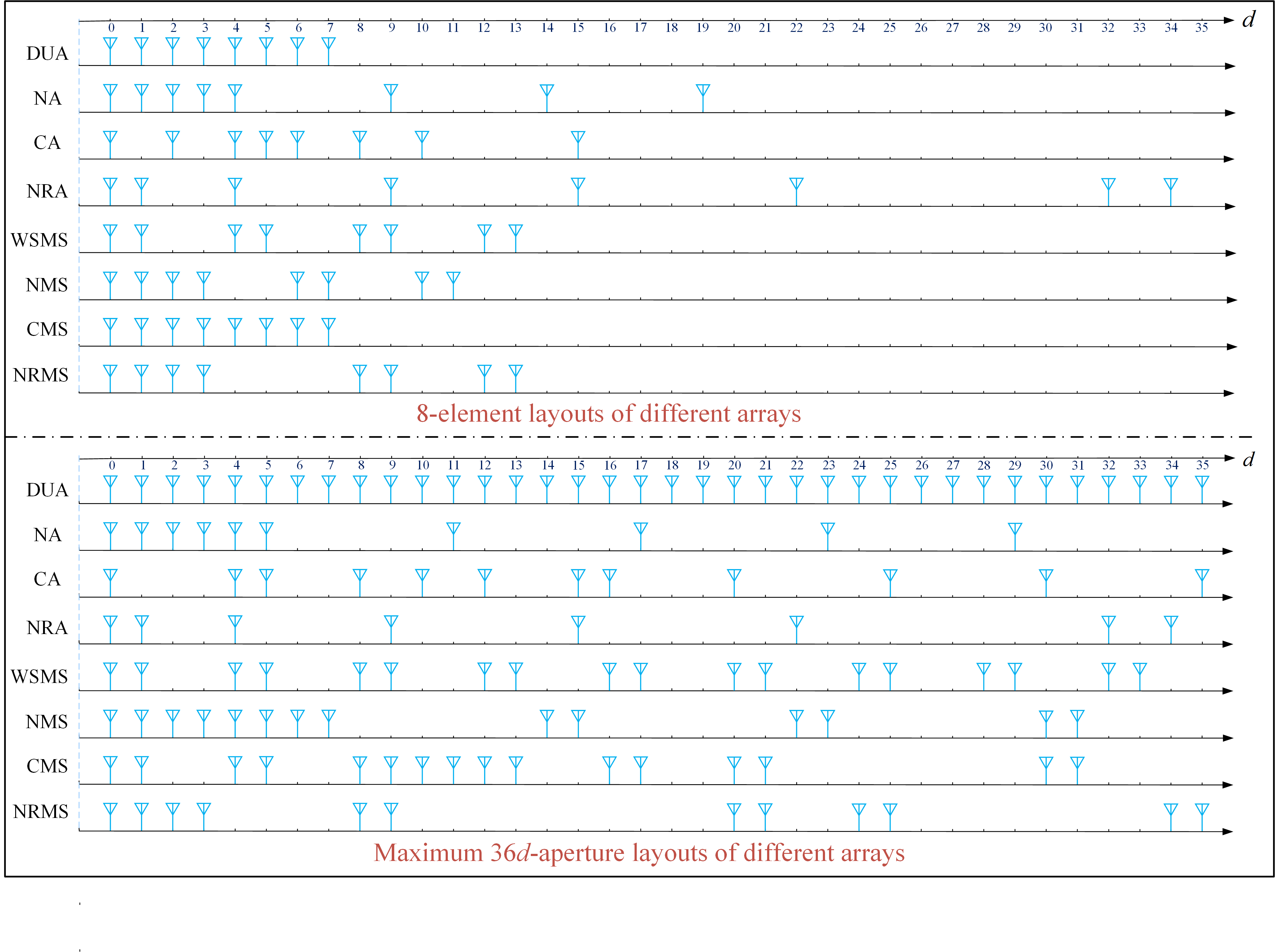}
	\caption{Antena positions of different arrays for two cases: (a) fixed number of atennas $8$, and (b) fixed maximum aperture $36d$. }\label{LAYOUT} 
\end{figure*} 
\subsection{What are SAs?}\label{D-SAs}
Generally, SAs can be defined in two ways: 
\begin{itemize}
	\item As a carefully pre-designed antenna layout with a small number of elements, it strategically achieves specific objectives like maximizing DoFs or minimizing redundancy.
	\item As a dynamic array that minimizes the number of active antennas and optimizing the antenna positions/weights while preserving system performance. This is also referred to as sparse array synthesis \cite{SAS}.
\end{itemize}
  These two definitions reflect different approaches to harnessing the advantages of SAs, each catering to distinct scenarios and optimization goals. Here are some typical SAs.
 \subsubsection{Coprime Arrays (CAs)} 
The CA is one of the first sparse arrays introduced with closed-form expressions for antenna positions. This means that the antenna locations or the SA configuration can be immediately obtained once the number of elements is given, without the need for any exhaustive search mechanisms. An $N$-element CA consists of a $Q$-element array with $Pd$ inter-element spacing and a $2P-1$-element array with $Qd$ inter-element spacing, where $P$ and $Q$ are coprime integers such that $P$ is smaller than $Q$, and $d$ denotes the half-wavelength inter-element spacing.

\subsubsection{Nested Arrays (NAs)}\label{NA}
The NA offer another closed-form expression for antenna positions, providing hole-free difference co-arrays, an improvement over CAs and an alternative to minimum redundancy arrays. An $N$-element two-level NA (assuming even $N$) comprises two $\frac{N}{2}$-element arrays with inter-element spacings of $d$ and $(\frac{N}{2}+1)d$, respectively. Particularly, the second array starts from $\frac{Nd}{2}$.

  \subsubsection{Non-Redundant Arrays (NRAs)}  
   Design of NRAs has been a topic with great
  interest due to their ability to provide the highest number of DoFs.   Due to the variety of non-redundant arrays, this article just considers one of them \cite{NRA}.

\subsection{Why Opt for SAs in XL-arrays?}

To cater to various S\&C applications with XL-arrays and overcome the hardware and algorithm challenges, efficient changes to array layouts are worthy of attention to enhance near-field S\&C.
We will compare the advantages of SAs over DUAs through five key features. 

\subsubsection{System Cost, Versatility and Scalability}  
SAs with fewer antennas save costs and energy, simplifying algorithms. They excel in precise beamforming and complex electromagnetic environments due to lower mutual coupling from sparse element distribution. Compared to DUAs, SAs offer greater versatility, customization, and scalability to larger arrays with reduced mutual coupling and minimal performance degradation, overcoming challenges faced by DUAs.
 
 \subsubsection{Sensing Ability}
SAs have been demonstrated to be highly effective in improving spatial resolution and providing a significantly higher number of sensing DoFs compared to DUAs.
The enhanced spatial resolution of SAs allows for more accurate localization and tracking of targets, making them well-suited for various applications. Additionally, the increased sensing DoFs provided by SAs enable better discrimination and characterization of complex electromagnetic environments, offering improved situational awareness and information gathering capabilities.
\subsubsection{Communication Ability}
Beyond sensing applications, SAs have proven valuable in wireless communications as well. Leveraging their high spatial resolution and increased DoFs, SAs can outperform DUAs in SDMA scenarios. Moreover, SAs' adaptability allows for tailoring beam patterns and coverage to suit specific communication scenarios by optimizing antenna positions, providing greater versatility in dynamically optimizing communication performances.

\subsection{How to Integrate SAs into XL-Arrays?}

Incorporating the introduced SAs into mmWave/THz XL-MIMO systems presents a challenge due to the significantly increased array aperture resulting from the growing number of antennas. For instance, let us consider a $128$-element two-level NA as described in Section \ref{NA}. It will incur an aperture of $4120d$ that is too large for practical deployment.
To inherit the advantages of SAs and account for the required number of antennas in high-frequency communications to overcome path loss challenges, we aim to design more efficient arrays for XL-arrays.
 
The array of subarray, also called as the partially-connected hybrid beamforming structure, show promise for practical mmWave/THz systems as it reduces hardware complexity by directly connecting each radio frequency (RF) chain to an array antenna. Building upon this concept, the widely-spaced multi-subarray (WSMS) structure was proposed in \cite{WSMS1}, where the inter-subarray spacing was increased to exploit the multiplexing benefit of the SW model. 

Recognizing that the equidistant inter-subarray spacing in WSMSs leads to spatial redundancy, we combine the property of SAs to present the concept of sparse subarrays. Additionally, we explore the typical SAs, giving rise to multi-subarrays (CMSs), nested multi-subarrays (NMSs), and non-redundant multi-subarrays (NRMSs). These sparse subarrays can be designed by treating each subarray as an antenna and following the rule of SA design.


Overall, Table \ref{TAB} presents the antenna positions and DoFs of different arrays in two cases. In the first case, all arrays consist of 8 antennas, revealing that SAs possess a larger number of DoFs than DUAs. Notably, NRAs exhibit the highest DoFs, but their large array aperture may not be practical for XL-systems. On the other hand, our proposed NRMSs offer a considerable number of DoFs with a more manageable aperture size. For the second case when fixing a maximum aperture of $36d$ for all arrays, SAs have significantly fewer antennas than DUAs, while maintaining a comparable number of DoFs. Interestingly, our proposed SAs outperform traditional SAs in terms of DoFs, and this DoF gap will widen as the aperture size increases.
  Furthermore,
 Fig. \ref{LAYOUT} visually illustrates the physical layout of the arrays for both cases in Table \ref{TAB}.
 \begin{figure*}
 	\centering 
 	\includegraphics[width=6.1in]{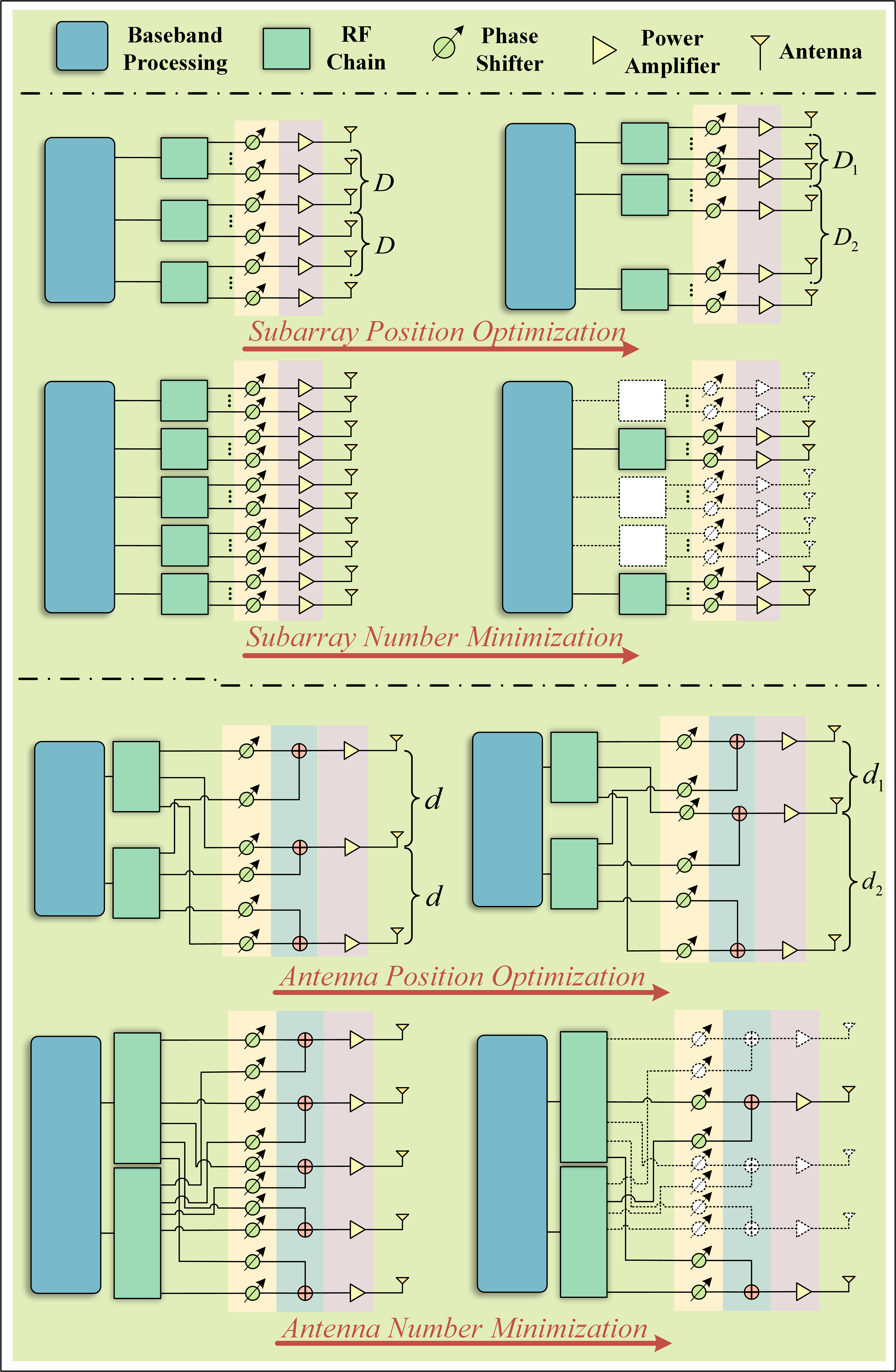}
 	\caption{ Four hybrid beamforming cases for XL-systems. The top four structures and the bottom four structures are partially-connected and fully-connected hybrid beamforming structures, respectively.
 	 }\label{HB} 
 \end{figure*} 
\subsection{How Will XL-SAs Function?}\label{APPL}
Having examined the advantages of SAs and the design considerations for XL-SAs, we now delve into potential applications where XL-SAs demonstrate promising prospects.

\subsubsection{Dynamic Near-Field Hybrid Beamforming}
With the flexibility offered by XL-SA, near-field hybrid beamforming becomes more versatile. Fig. \ref{HB} illustrates four cases of hybrid beamforming optimized with SAs. For partially-connected structures, optimizing the subarray positions becomes crucial when maximizing near-field effects under a certain subarray count. Reducing the number of subarrays to improve energy efficiency while maintaining considerable system performance is also significant. For fully-connected structures, these scenarios can be adapted from partially-connected structures, with the difference lying in the optimization of antenna elements rather than subarrays.

\subsubsection{Active XL-RIS}

Active RISs offer significant advantages over passive RIS in terms of channel capacity. Moreover, active RIS facilitates the adoption of more efficient channel estimation methods, enabling accurate and timely estimation of channel conditions \cite{S-RIS}. This can potentially resolve the issue of significant pilot overhead required for estimating near-field XL-RIS channels \cite{XL-RIS}.
 Nevertheless, the high hardware cost and energy consumption limit the practicality of fully-active RISs in XL-systems. To address this, active XL-RISs are proposed, wherein a small subset of elements are active while the rest remain passive. Two approaches are considered: pre-designed SA patterns (NMSs, CMSs, and NRMSs) and dynamic deployment using electronic switches to adapt to different scenarios. 

\subsubsection{Enhanced SDMA}
XL-SAs show greater potential for near-field SDMA compared to XL-DUAs, attributed to increased degrees of freedom and spatial resolutions. XL-DUAs may face limitations when multiple closely located users access the system, leading to performance errors. Additionally, thier range resolution decreases as the angle between the access user and the array normal becomes larger \cite{NCE}. To overcome this challenge, XL-SAs with flexible array layouts optimize antenna positions, improving system performance.
 
\subsubsection{Enhanced Near-Field ISAC}

As mentioned earlier, near-field ISAC shows tremendous potential. Similar to existing sensing systems, future ISAC systems will demand sparse arrays with high DoFs and high resolution to enable advanced sensing applications. Moreover, this will further improve the capabilities of near-field communication, making the integration of S\&C in the near-field region even more promising. Current far-field or near-field ISAC has limited DoFs for an all-encompassing solution, which can be augmented by SAs optimizing array layout to enhance ISAC from hardware to algorithm.

\subsubsection{Near-Field Pattern-Reconfigurable MIMO (PR-MIMO)}

PR-MIMO is developed to overcome the performance limitations of aperture-restricted MIMO systems. Unlike traditional MIMO with fixed antenna patterns, PR-MIMO enables dynamic pattern adjustment for each antenna, allowing high-directional and adaptive beamforming. However, when using XL-arrays, PR-MIMO faces challenges: high hardware costs due to complex circuitry and feeding, and increased pattern optimization complexity. Therefore, XL-SAs are better suited for PR-MIMO in XL-array configurations. They provide enhanced performance while reducing the hardware cost and pattern optimization complexity, making them a more suitable choice for implementing PR-MIMO in XL-systems.

\subsubsection{Enhanced MIMO-LoS Communications}
MIMO-LoS plays a critical role in high-frequency communications, maritime communications, and unmanned aerial vehicle communications. When utilizing XL-arrays, the SW channel enhances MIMO-LoS, leading to improved multiplexing and capacity. XL-SAs with their flexible and larger array apertures outperform XL-DUAs in MIMO-LoS communication, even with the same number of antennas. Additionally, optimizing antenna positions can further achieve optimal MIMO-LoS transmission.
\section{Numerical Results}

To evaluate the potential of XL-SAs in near-field S\&C applications, we conduct two simulations comparing their performance against XL-DUAs. Practical aperture configurations are considered, excluding NAs, CAs, and NRAs as benchmarks. The simulations are performed at a system frequency of $100$ GHz, with $512$ antennas, and a received signal-to-noise ratio of $0$ dB. The multi-subarray structure comprises $8$ subarrays, each containing $64$ antennas, is considered.

\subsection{Sensing Performance}
\begin{figure}
	\centering 
	\includegraphics[width=3.22in]{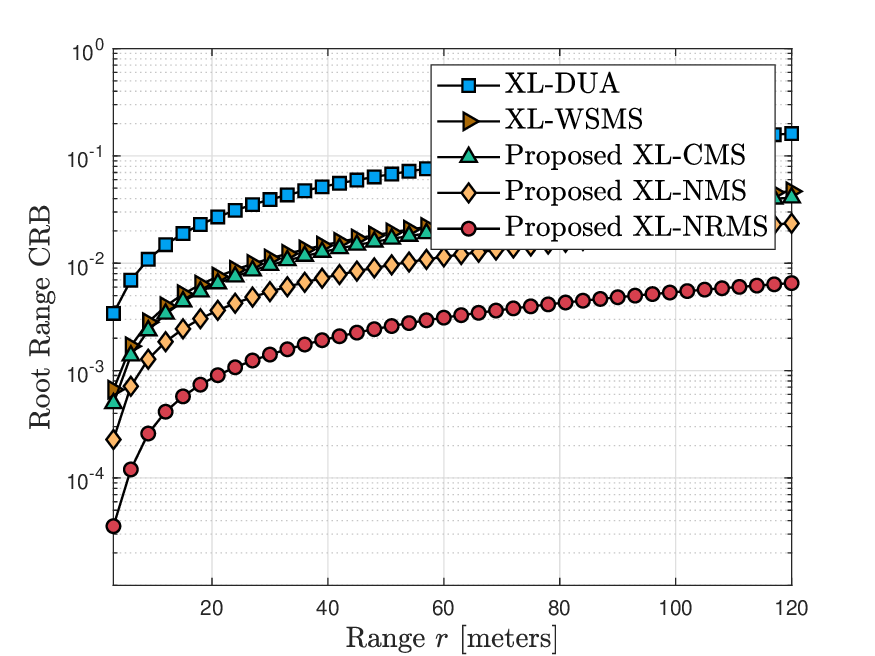}
	\caption{The root range CRB versus the range $r$.}\label{Sensing} 
\end{figure} 

From Fig. \ref{Sensing}, which displays the root range CRB performance versus the range $r$, it is evident that our proposed XL-SAs outperform XL-DUAs in terms of the root range CRB. Among these arrays, XL-NRMSs exhibit the best performance, showcasing their high spatial resolution capabilities. Furthermore, the root range CRB increases with the range, indicating that range estimation can be more precise in the near-field region. Alternatively, when sensing a far-field object located, for example, $120$ meters away, XL-DUAs are limited due to the high estimation error. In this sense, XL-SAs offer an advantage by providing enhanced performance for robust range sensing in such scenarios. 
\subsection{Communication Performance} 
\begin{figure}
		\centering
		\includegraphics[width=3.22in]{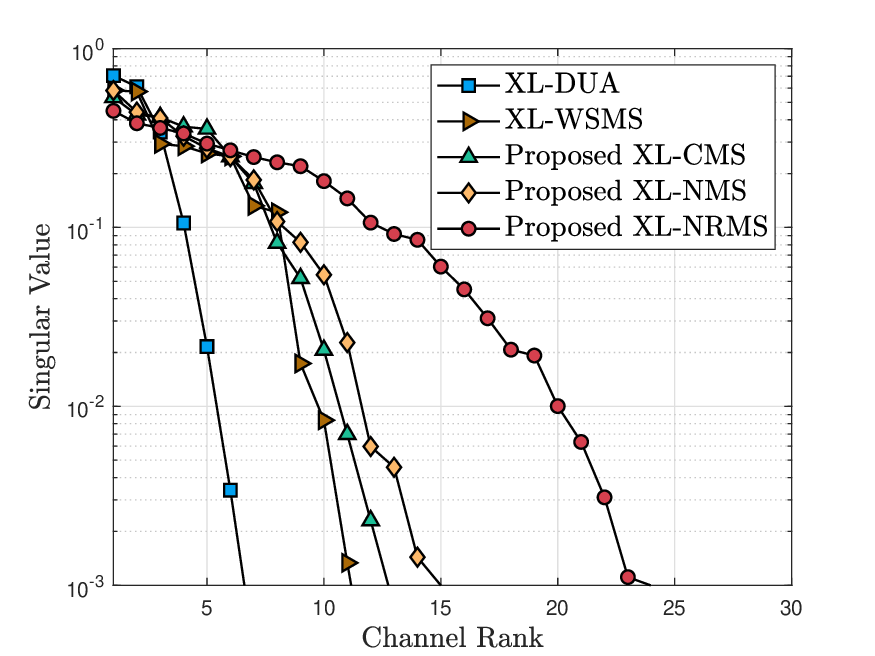}
		\caption{The singular value versus the channel rank.} \label{Comm}  
\end{figure}

In this evaluation, we analyze the channel rank as a communication performance indicator. For the MIMO case, we consider equal numbers of antennas for both the transmitter and the receiver, with a fixed distance of $100$ meters between them. We use singular value decomposition to deal with the SW channel. We define singular values greater than $10^{-3}$ as valid channel paths, and the channel rank is then determined by counting the number of such singular values.
As shown in Fig. \ref{Comm}, the XL-DUA has a maximum channel rank of $6$, allowing support for only $6$ simultaneous data streams. In contrast, our proposed XL-SAs provide more DoFs for spatial multiplexing, with XL-NRMSs demonstrating the strongest multiplexing capability. 
\section{Open Challenges and Directions} 
To support those applications introduced in Section \ref{APPL} and fully exploit the potential of XL-SAs, certain challenges need to be addressed seriously.

\subsection{SA View}

\subsubsection{SA Design}

This article introduces promising XL-SAs utilizing a multi-subarray structure, effectively enhancing their S\&C performances. Successful implementation paves the way for integrating more effective SAs into XL-systems, motivating future research in this area for improved S\&C performance across various applications.
\subsubsection{SA Optimization}

Developing real-time algorithms to dynamically optimize XL-SAs' antenna positions and numbers presents significant challenges.  Exploring a joint optimization framework for XL-SAs and other techniques, including near-field hybrid beamforming, holds substantial potential. It offers solutions for various demands, such as maximizing energy efficiency and addressing diverse requirements in wireless communication systems.

\subsection{S\&C View}
\subsubsection{Wideband Transmission} 
In ultra-wideband XL-systems, beam squint and near-field effects pose challenges for beam management, channel estimation, and hybrid beamforming. XL-SAs offer potential for decreased computational complexity with fewer antennas. However, algorithm design must leverage SA properties to address spatial challenges and maximize XL-SAs' benefits in diverse applications.
\subsubsection{Channel Estimation} 

Channel estimation for XL-SAs presents diverse challenges in various scenarios. Understanding the SW beam squint effect in wideband systems is crucial for accurate beamspace channel reconstruction \cite{NCE3,NCE5}, but XL-SAs may exhibit distinct beam squint trajectories compared to XL-DUAs. In MIMO scenarios, an efficient joint NLoS and LoS recovery method should be investigated for XL-SAs, considering the significant difference between the array manifolds of the LoS and NLoS. In XL-RIS applications, active XL-RISs with XL-SAs can greatly aid channel estimation through simultaneous reception and reflection exploitation.

\subsubsection{Sensing and Tracking}

Extensive research on near-field sensing with DUAs has been conducted. The introduction of XL-SAs opens up opportunities to enhance sensing applications, transitioning from far-field to near-field SA sensing. Furthermore, emphasizing the significance of near-field parameter tracking, despite its limited attention, is crucial. The extra range parameter in the near-field region necessitates meticulous design of tracking methods for XL-SAs to ensure optimal utilization in near-field parameter tracing applications.
\subsubsection{Dynamic S\&C and Itegration}
The ultimate goal of XL-SAs is to achieve diversified S\&C objectives: improving energy efficiency, expanding near-field coverage, and optimizing trade-offs. First, efficient SA optimization algorithms enable XL-SAs to reduce power consumption while maintaining system performance. Compared to XL-DUAs, XL-SAs demonstrate significantly enhanced characteristics in SW properties when given a fixed number of antennas. Moreover, dynamic antenna positions provide ISAC with more DoFs for optimizing the S\&C trade-off at a system level.

 Achieving these goals involves developing novel algorithms and efficient system designs, presenting new opportunities for future research in XL-SAs.

\section{Conclusions}\label{Con}
In this article, we propose XL-SAs as a means to enhance near-field S\&C. Toward this end, we start by exploring near-field XL-arrays using the SW model, covering their hardware/algorithm challenges, fundamental potentials, and promising applications. The findings suggest that near-field S\&C will drive a new trend in 6G technologies, while also  highlighting the existence of opportunities for further advancements in this field. Following a concise overview of SAs, we investigate the rationale for choosing SAs and the approach involved in their seamless integration into XL-arrays. After that, we highlight our proposed XL-SAs along with numerical results in terms of CRBs and multiplexing, demonstrating the significant potential of XL-SAs for enhancing near-field S\&C. Beyond this, the article discusses prospective applications, key challenges, and directions of
future research related to near-field S\&C enpowered by XL-SAs, underscoring their pivotal significance in the forthcoming 6G era.

\bibliographystyle{IEEEtran}
\bibliography{reference.bib}

\vspace{12pt}

\end{document}